\documentclass{article}
\usepackage[english]{babel}
\usepackage[letterpaper,top=2cm,bottom=2cm,left=3cm,right=3cm,marginparwidth=1.75cm]{geometry}

\usepackage{amsmath}
\usepackage{graphicx}
\usepackage{cite}
\usepackage{color}
\usepackage[colorlinks=true, allcolors=blue]{hyperref}

\definecolor{green}{rgb}{0.1,0.5,0.0}

\usepackage{ifthen} 
\newboolean{uprightparticles}
\setboolean{uprightparticles}{false} 

\usepackage{xspace} 
\usepackage{upgreek}

\def\MagUp {\mbox{\em Mag\kern -0.05em Up}\xspace}

\ifthenelse{\boolean{uprightparticles}}%
{

 \def\Ppi         {\ensuremath{\uppi}\xspace}

 \def\Pphi        {\ensuremath{\upphi}\xspace}

 \def\PDelta      {\ensuremath{\Delta}\xspace}                 
 \def\PXi      {\ensuremath{\Xi}\xspace}                 
 \def\PLambda      {\ensuremath{\Lambda}\xspace}                 
 \def\PSigma      {\ensuremath{\Sigma}\xspace}                 
 \def\POmega      {\ensuremath{\Omega}\xspace}                 
 \def\PUpsilon      {\ensuremath{\Upsilon}\xspace}                 
 
 \def\PB      {\ensuremath{\mathrm{B}}\xspace}                 
                  
 \def\PD      {\ensuremath{\mathrm{D}}\xspace}

 \def\PK      {\ensuremath{\mathrm{K}}\xspace}

 \def\Pi      {\ensuremath{\mathrm{i}}\xspace}

 \def\Ps      {\ensuremath{\mathrm{s}}\xspace}

}
{

 \def\Ppi         {\ensuremath{\pi}\xspace}

 \def\Pphi        {\ensuremath{\phi}\xspace}

 \mathchardef\PDelta="7101
 \mathchardef\PXi="7104
 \mathchardef\PLambda="7103
 \mathchardef\PSigma="7106
 \mathchardef\POmega="710A
 \mathchardef\PUpsilon="7107
                  
 \def\PB      {\ensuremath{B}\xspace}                 
                  
 \def\PD      {\ensuremath{D}\xspace}

 \def\PK      {\ensuremath{K}\xspace}

 \def\Pi      {\ensuremath{i}\xspace}

 \def\Ps      {\ensuremath{s}\xspace}

}

\makeatletter
\ifcase \@ptsize \relax
  \newcommand{\miniscule}{\@setfontsize\miniscule{4}{5}}
\or
  \newcommand{\miniscule}{\@setfontsize\miniscule{5}{6}}
\or
  \newcommand{\miniscule}{\@setfontsize\miniscule{5}{6}}
\fi
\makeatother

\DeclareRobustCommand{\optbar}[1]{\shortstack{{\miniscule (\rule[.5ex]{1.25em}{.18mm})}
  \\ [-.7ex] $#1$}}

\def\squark    {{\ensuremath{\Ps}}\xspace}

\def\pion   {{\ensuremath{\Ppi}}\xspace}

\def\pim    {{\ensuremath{\pion^-}}\xspace}
\def\pipm   {{\ensuremath{\pion^\pm}}\xspace}
\def\pimp   {{\ensuremath{\pion^\mp}}\xspace}

\def\kaon    {{\ensuremath{\PK}}\xspace}
  \def\Kbar    {{\kern 0.2em\overline{\kern -0.2em \PK}{}}\xspace}

\def\KorKbar    {\kern 0.18em\optbar{\kern -0.18em K}{}\xspace}

\def\Kp      {{\ensuremath{\kaon^+}}\xspace}
\def\Km      {{\ensuremath{\kaon^-}}\xspace}
\def\Kpm     {{\ensuremath{\kaon^\pm}}\xspace}
\def\Kmp     {{\ensuremath{\kaon^\mp}}\xspace}

  \def\Dbar    {{\kern 0.2em\overline{\kern -0.2em \PD}{}}\xspace}
\def\D       {{\ensuremath{\PD}}\xspace}

\def\DorDbar    {\kern 0.18em\optbar{\kern -0.18em D}{}\xspace}

\def\Dpm     {{\ensuremath{\D^\pm}}\xspace}

\def\Dsp     {{\ensuremath{\D^+_\squark}}\xspace}

\def\Dspm    {{\ensuremath{\D^{\pm}_\squark}}\xspace}

\def\B       {{\ensuremath{\PB}}\xspace}
\def\Bbar    {{\ensuremath{\kern 0.18em\overline{\kern -0.18em \PB}{}}}\xspace}

\def\BorBbar    {\kern 0.18em\optbar{\kern -0.18em B}{}\xspace}

\def\Bs      {{\ensuremath{\B^0_\squark}}\xspace}
\def\BsorBsbar    {\kern 0.18em\optbar{\kern -0.18em \Bs}{}\xspace}
\def\Bsb     {{\ensuremath{\Bbar{}^0_\squark}}\xspace}

  \def\Y#1S{\ensuremath{\PUpsilon{(#1S)}}\xspace}

\def\Lbar        {{\ensuremath{\kern 0.1em\overline{\kern -0.1em\PLambda}}}\xspace}
\def\LorLbar    {\kern 0.18em\optbar{\kern -0.18em \PLambda}{}\xspace}

\def\to                 {\ensuremath{\rightarrow}\xspace}

\def\AT#1     {\ensuremath{A_{\mathrm{T}}^{#1}}\xspace}           

\def\C#1      {\ensuremath{\mathcal{C}_{#1}}\xspace}                       
\def\Cp#1     {\ensuremath{\mathcal{C}_{#1}^{'}}\xspace}                    
\def\Ceff#1   {\ensuremath{\mathcal{C}_{#1}^{\mathrm{(eff)}}}\xspace}        
\def\Cpeff#1  {\ensuremath{\mathcal{C}_{#1}^{'\mathrm{(eff)}}}\xspace}       
\def\Ope#1    {\ensuremath{\mathcal{O}_{#1}}\xspace}                       
\def\Opep#1   {\ensuremath{\mathcal{O}_{#1}^{'}}\xspace}                    

\newcommand{\tev}{\ifthenelse{\boolean{inbibliography}}{\ensuremath{~T\kern -0.05em eV}}{\ensuremath{\mathrm{\,Te\kern -0.1em V}}}\xspace}
\newcommand{\gev}{\ensuremath{\mathrm{\,Ge\kern -0.1em V}}\xspace}
\newcommand{\mev}{\ensuremath{\mathrm{\,Me\kern -0.1em V}}\xspace}
\newcommand{\kev}{\ensuremath{\mathrm{\,ke\kern -0.1em V}}\xspace}
\newcommand{\ev}{\ensuremath{\mathrm{\,e\kern -0.1em V}}\xspace}
\newcommand{\gevc}{\ensuremath{{\mathrm{\,Ge\kern -0.1em V\!/}c}}\xspace}
\newcommand{\mevc}{\ensuremath{{\mathrm{\,Me\kern -0.1em V\!/}c}}\xspace}
\newcommand{\gevcc}{\ensuremath{{\mathrm{\,Ge\kern -0.1em V\!/}c^2}}\xspace}
\newcommand{\gevgevcccc}{\ensuremath{{\mathrm{\,Ge\kern -0.1em V^2\!/}c^4}}\xspace}
\newcommand{\mevcc}{\ensuremath{{\mathrm{\,Me\kern -0.1em V\!/}c^2}}\xspace}

\def\invfb   {\ensuremath{\mbox{\,fb}^{-1}}\xspace}

\def\gsim{{~\raise.15em\hbox{$>$}\kern-.85em
          \lower.35em\hbox{$\sim$}~}\xspace}
\def\lsim{{~\raise.15em\hbox{$<$}\kern-.85em
          \lower.35em\hbox{$\sim$}~}\xspace}

\def\tell1  {TELL1\xspace}
\def\ukl1   {UKL1\xspace}

\begin{document}

\begin{flushright}  
{\small  
 SI-HEP-2021-030
}  
\end{flushright}  
\begin{center}
{\Large \bf \boldmath Testing the Standard Model with $CP$-asymmetries \\[1mm] in flavour-specific non-leptonic decays} \\[7mm]
{\large Tim Gershon$^a$, Alexander Lenz$^b$,  Aleksey~V.~Rusov$^b$ and Nicola Skidmore$^c$}
\\[4mm]
$^a$ 
{\it Department of Physics, University of Warwick, Coventry, CV4 7AL, UK}
\\[1mm]
$^b$
{\it Physik Department, Universit\"at Siegen, Walter-Flex-Str. 3, 57068 Siegen, Germany} \\[1mm]
$^c$
{\it University of Manchester, Schuster Building, Manchester, M13 9PL, UK}
\end{center}

\begin{abstract}
Motivated by recent indications that the rates of colour-allowed non-leptonic channels are not in agreement with their Standard Model expectations based on QCD factorisation, we investigate the potential to study CP asymmetries with these decays.
In the Standard Model, these flavour-specific decays are sensitive to CP violation in $B^0_{(s)}$--$\bar{B}^0_{(s)}$ mixing, which is predicted with low uncertainties and can be measured precisely with semileptonic decays.
If there are beyond Standard Model contributions to the non-leptonic decay amplitudes, there could be significant enhancements to the CP asymmetries.
Measurements of these quantities therefore have potential to  identify BSM effects without relying on Standard Model predictions that might be affected by hadronic effects.
We discuss the experimental prospects, and note the excellent potential for a precise determination of the CP asymmetry in $\bar{B}_s \to D_s^+ \pi^-$ decays by the LHCb experiment.  
\end{abstract}

\section{Introduction}

Recent theoretical investigations~\cite{Huber:2016xod,Bordone:2020gao,Cai:2021mlt,Endo:2021ifc} have revealed a discrepancy between experimental measurements of the rates of colour-allowed non-leptonic decays~\cite{LHCb:2012wdi,LHCb:2013vfg,LHCb:2021qbv} and their predicted values in the Standard Model (SM), based on QCD factorisation~\cite{Beneke:2000ry}.
While the origin of this disagreement could be due to unaccounted-for QCD effects or maybe partly due to ultra-soft photon effects
\cite{Beneke:2021jhp}, there is also an enticing possibility that physics beyond the Standard Model may be contributing.  
It is therefore of interest to investigate theoretically clean observables that could help to address this possibility.
As we will show, the CP asymmetry in the flavour-specific decay $\bar{B}_s \to D_s^+ \pi^-$ is well suited for this purpose.

We denote the decay amplitude describing the transition of the flavour eigenstate $B_q$ ($ q = d,s$) to the final state~$f$ by ${\cal A}_f$; 
for the decay of a $\bar{B}_q$ eigenstate into 
$f$ we use the notation $\bar{\cal A}_f$.
The underlying flavour changing weak quark transitions are described by the effective Hamiltonian. Thus we can write:
\begin{equation}
{\cal A}_f = \langle f | {\cal H}_{\rm eff} |B_q \rangle \; ,
\hspace{1cm}
\bar{\cal A}_f = \langle f | {\cal H}_{\rm eff} | \bar{B}_q \rangle \; ,
\label{amplitude}
\end{equation}
with obvious extension to the notation for decays into the CP conjugate final states $\bar{f}$.
A {\it flavour-specific decay} of the $B_q$ meson is defined by the condition, see e.g. \cite{osti_1151538,Artuso:2015swg}
\begin{itemize}\setlength{\itemindent}{10mm}
\item [\bf C1:]    $ {\cal A}_{\bar{f}} = 0 = \bar{\cal A}_f  $.  
\end{itemize}
This condition states that the  meson $B_q$ cannot decay into the CP conjugate final state $\bar{f}$, and $\bar{B}_q$ cannot decay into $f$.
Examples of $\bar{B}_s$ decays that are flavour-specific in the SM include semileptonic decays such as $\bar{B}_s \to D_s^+\ell^- \bar \nu_{\ell}$, and non-leptonic decays such as $\bar{B}_s \to D_s^+\pi^-$ and $\bar{B}_s \to K^+\pi^-$.
There are corresponding flavour-specific $\bar{B}^0$ decays to the $D^+\ell^- \bar \nu_{\ell}$, $D^+K^-$ and $K^-\pi^+$ final states.

Demanding further the absence of direct  CP violation in the decay $B_q \to f$ we get a second condition,
\begin{itemize}\setlength{\itemindent}{10mm}
\item [\bf C2:]     $ \bar{\cal A}_{\bar{f}}  = {\cal A}_f $.
\end{itemize}
Within the SM, the semileptonic decays and the non-leptonic decays $\bar{B}_s \to D_s^+ \pi^-$ and $\bar{B}^0 \to D^+ K^-$  are expected to have negligible direct CP violation, while by contrast the charmless non-leptonic decays $\bar{B}_s \to K^+\pi^-$ and $\bar{B}^0 \to K^-\pi^+$ do not satisfy condition {\bf C2}~\cite{Belle:2012dmz,BaBar:2012fgk,LHCb:2020byh}.

Due to weak interactions, transitions like $\bar{B}_q \leftrightarrow B_q$ are possible
via box diagrams and we define the meson mass eigenstates 
$|B_{q,H} \rangle$ ($H$ = heavy, mass $M_{H}^q $ and decay rate $\Gamma_H^s$)
and $|B_{q,L} \rangle$ ($L$ = light, mass $M_{L}^q $ and decay rate $\Gamma_L^s$) as linear combinations of the flavour eigenstates:
\begin{eqnarray}
| B_{q,L} \rangle & = & p |B_q  \rangle+ q |\bar{B}_q  \rangle\; ,
\\
| B_{q,H} \rangle & = & p |B_q  \rangle- q |\bar{B}_q  \rangle\; ,
\end{eqnarray}
with $|p|^2 + |q|^2 = 1$.
The ratio of the magnitudes of the coefficients $p$ and $q$, as well as the mass difference 
$\Delta M_q = M_{H}^q - M_{L}^q $
and the decay rate difference $\Delta \Gamma_q =  \Gamma_L^q - \Gamma_H^q$ can
be expressed in terms of the absorptive part $\Gamma_{12}^q$ and the dispersive part $M_{12}^q$ of the box  diagrams,
\begin{eqnarray}
\Delta M_q \approx 2 |M_{12}^q| \, ,
&&
\Delta \Gamma_q \approx 2 |\Gamma_{12}^q| \cos \phi_{12}^q\, , 
\\
\left| \frac{q}{p} \right| \approx
1 - \frac{a_{\rm fs}^q}{2}
      \, ,
&&
a_{\rm fs}^q \approx \frac{|\Gamma_{12}^q|  }{|M_{12}^q|  } \sin   \phi_{12}^q \; ,
\label{abbrev}
\end{eqnarray}
with $\phi_{12}^q = \arg (- M_{12}^q/\Gamma_{12}^q)$.

To measure the $a_{\rm fs}^q$ parameter, which quantifies CP violation in mixing, it is necessary to study neutral mesons that mix before decaying. 
The general time evolution of the decay rate of neutral $B_q$ mesons, which decay with flavour opposite to that at production, is given 
(see e.g.\ \cite{osti_1151538,Artuso:2015swg}) by
\begin{eqnarray}
\Gamma \left[\bar{B}_q (t) \to f \, \right] 
&  = &  
N_f \left|{\cal A}_f \right|^2
\frac{\left(1+|\lambda_f|^2 \right)}{2} (1+{a}_{\rm fs}^q) \, e^{-\Gamma_q t}
\biggl\{\cosh \left( \frac{\Delta \Gamma_q t}{2} \right)
- \frac{1-|\lambda_f|^2 }{1+|\lambda_f|^2 } \cos \left( \Delta M_q t\right)
\nonumber \\[2mm] 
& &
\hspace*{30mm}
- \, \frac{2 \, {\rm Re} (\lambda_f)}{1+|\lambda_f|^2}
  \sinh \left( \frac{\Delta \Gamma_q t }{2} \right)
+ \frac{2 \, {\rm Im} (\lambda_f )}{1+|\lambda_f|^2 }
  \sin  \left( \Delta M_q t\right)
\biggr\}
 \; ,
\label{GammabarBf}
\\[2mm]
\Gamma \left[B_q (t) \to \bar{f} \, \right] 
& = & 
N_f \left|\bar{\cal A}_{\bar f}\right|^2
\frac{\left(1+|\lambda_{\bar{f}}|^{-2}\right)}{2} (1-{a}_{\rm fs}^q) \,
e^{-\Gamma_q t}
\biggl\{\cosh \left( \frac{\Delta \Gamma_q t}{2} \right)
- \frac{1-|\lambda_{\bar{f}}|^{-2} }{1+|\lambda_{\bar{f}}|^{-2} }
\cos  \left( \Delta M_q t\right)
\nonumber 
\\[2mm]
& &
\hspace*{30mm}
- \, \frac{2 \, {\rm Re} (\frac{1}{\lambda_{\bar{f}}} )}{1+|\lambda_{\bar{f}}|^{-2} }
  \sinh \left( \frac{\Delta \Gamma_q t }{2} \right)
+ \frac{2 \, {\rm Im} (\frac{1}{\lambda_{\bar{f}}} )}{1+|\lambda_{\bar{f}}|^{-2} }
  \sin  \left( \Delta M_q t\right)
\biggr\}
\label{GammaBbarf} \; .
\end{eqnarray}
Here $\Gamma_q = (\Gamma_L^q + \Gamma_H^q)/2$, 
$N_f$ encodes a time-independent normalisation factor, including phase space effects, 
and the quantities $\lambda_f$ and $\lambda_{\bar{f}}$ are defined as
\begin{equation}
\lambda_f = \frac{q}{p} \frac{\bar{\cal A}_f}{{\cal A}_f} \quad \text{and} \quad
\lambda_{\bar{f}} = \frac{q}{p} \frac{\bar{\cal A}_{\bar{f}}}{{\cal A}_{\bar{f}}}  \; .
\label{lambdaf}
\end{equation}
In what follows, we will consider the flavour-specific CP asymmetry 
(often called semileptonic CP asymmetry), defined as 
\begin{equation}
A_{\rm  fs}^q = 
\frac{\Gamma \left(\bar{B}_q (t)  \to f\right) - \Gamma \left({B}_q (t)  \to \bar{f}\right) }
{\Gamma \left(\bar{B}_q (t)  \to f \right) + \Gamma \left({B}_q (t)  \to \bar{f}\right) }\,.
\label{ea:def_asl}
\end{equation}

\section{\boldmath $ A_{\rm  fs}^q $ within the SM}
Within the SM we get for flavour-specific decays due to condition {\bf C1}:
$\lambda_f = 0 = 1/ \lambda_{\bar{f}}$, 
the simplified time evolution
\begin{eqnarray}
\Gamma \left[\bar{B}_q (t) \to f \, \right] 
& = &  
\frac{1}{2} N_f \left|{\cal A}_f\right|^2
(1 + {a}_{\rm fs}^q) \, e^{-\Gamma_q t}
\label{GammabarBf2}
X_q^-(t)
\; ,
\\
\Gamma \left[B_q (t) \to \bar{f} \, \right] 
& = & 
\frac{1}{2} N_f \left|\bar{\cal A}_{\bar f}\right|^2
(1-{a}_{\rm fs}^q) \, e^{-\Gamma_q t}
\label{GammaBbarf2}
X_q^-(t)
 \; , 
\end{eqnarray}
with the short-hand notation
\begin{eqnarray}
X_q^\pm (t)& \equiv & \cosh \left( \frac{\Delta \Gamma_q \, t }{2} \right)
\pm \cos  \left( \Delta M_q t\right)\,.
\end{eqnarray}
This leads to
\begin{equation}
A_{\rm  fs}^q = 
\frac{ \left|{\cal A}_f\right|^2 (1+{a}_{\rm fs}^q) 
     - \left|\bar{\cal A}_{\bar f}\right|^2 (1-{a}_{\rm fs}^q) }
{ \left|{\cal A}_f\right|^2 (1+{a}_{\rm fs}^q) 
     + \left|\bar{\cal A}_{\bar f}\right|^2 (1-{a}_{\rm fs}^q) }
                   \, .
\label{ea:def_asl2}
\end{equation}
Note that this result for the asymmetry of time-dependent decay rates given in Eq.~\eqref{ea:def_asl} does not depend on time.
Condition {\bf C2} further gives $ \bar{\cal A}_{\bar{f}}  = {\cal A}_f $
and thus
\begin{equation}
A_{\rm  fs}^q = {a}_{\rm fs}^q                   \, .
\label{ea:def_asl3}
\end{equation}
The SM predictions for ${a}_{\rm fs}^q$ are tiny, so that measurements of ${a}_{\rm fs}^q$ are generally considered to be null tests of the SM.
Based on the calculations in Refs.~\cite{Beneke:1996gn,
Beneke:1998sy,
Beneke:2003az,
Ciuchini:2003ww,
Lenz:2006hd,
FermilabLattice:2016ipl,
Kirk:2017juj,
King:2019lal,
Dowdall:2019bea,
DiLuzio:2019jyq,
Davies:2019gnp,
Lenz:2020efu}, the most recent predictions~\cite{Lenz:2019lvd} are
\begin{eqnarray}
    a_{\rm fs}^d = (-4.73 \pm 0.42) \cdot 10^{-4}  \, ,
    && 
    a_{\rm fs}^s =  (2.06 \pm 0.18) \cdot 10^{-5}  \, ,
    \nonumber
    \\ 
    \left| \frac{\Gamma_{12}^d }{M_{12}^d } \right| =   (4.82 \pm 0.65) \cdot 10^{-3}  \, ,
    &&
    \left| \frac{\Gamma_{12}^s}{M_{12}^s } \right| =
       (4.82 \pm 0.64) \cdot 10^{-3}  \, ,
    \nonumber
    \\     
    \phi_{12}^d  = (-98\pm 19) \, {\rm mrad} 
    = (-5.6 \pm 1.1)^\circ  \, ,
    &&
    \phi_{12}^s  = (4.3 \pm  0.8) \, {\rm mrad} 
     = (0.25 \pm 0.05)^\circ    \, .
\end{eqnarray}
Measurements of ${a}_{\rm fs}^q$ have so far been made almost exclusively with semileptonic final states (motivating the alternative notation ${a}_{\rm sl}^q$).
The latest world averages~\cite{HFLAV:2019otj}, based mainly on the results of Refs.~\cite{BaBar:2014bbb,BaBar:2013xng,Belle:2005cou,D0:2012idx,D0:2012hft,LHCb:2014dcj,LHCb:2016ssr}, are
\begin{eqnarray}
     a_{\rm sl}^d = a_{\rm fs}^d =    (-21 \pm 17) \cdot 10^{-4}  \, ,
    && 
    a_{\rm sl}^s = a_{\rm fs}^s =  (-60 \pm 280) \cdot 10^{-5}     \, .
\label{eq:asl_HFLAV}
\end{eqnarray}
The experimental precision for these quantities is expected to increase considerably.
Refs.~\cite{LHCb:2018roe,Cerri:2018ypt} quote an estimated precision of $\pm \,2 \cdot 10^{-4}$ for $a_{\rm sl}^d$ and $\pm \,30 \cdot 10^{-5}$ for $a_{\rm sl}^s$, achievable by the LHCb experiment with an integrated luminosity of $300\,{\rm fb}^{-1}$.
While for $a_{\rm sl}^d$ this approaches the precision necessary to test the SM prediction, this large data sample will still not be sufficient to observe a non-zero value at the SM expectation of $a_{\rm fs}^s$.
Nevertheless, significantly more precise results than currently available will provide stringent constraints on beyond SM contributions to $\Gamma_{12}^s$ and $M_{12}^s$, as discussed below.
The possibility to determine these asymmetries with flavour-specific non-leptonic decays has not been considered widely, as the lower yields available would result in considerably larger uncertainties compared to the semileptonic decay.

\section{\boldmath $A_{\rm  fs}^q$ beyond the SM}
There are several possible ways that the quantities
$A_{\rm fs}^q $ could be modified in the presence of new physics.
We discuss these in turn below.

\subsection{\boldmath Modification of $M_{12}$}
\label{sec:Mod_M12}
General new physics effects in the dispersive part of $B$ mixing can be parameterised as (in the convention of \cite{Lenz:2006hd,Lenz:2011zz}) 
\begin{equation}
    M_{12}^q ~=~  M_{12}^{q, \rm SM} \cdot \Delta_q
    ~=~ \left| M_{12}^{q, \rm SM} \right| \cdot 
    \left| \Delta_q
    \right| e^{i (\phi_q^{M, \rm SM} + \phi_q^\Delta)} \, .
\end{equation}
The parameters $\left| \Delta_q\right|$ are constrained to be close to unity, with around $\pm 10 \%$ uncertainty, by the
agreement of the experimental measurements~\cite{HFLAV:2019otj,LHCb:2016gsk,LHCb:2021moh} of the mass differences with the theoretical determinations via 
$ \Delta M_q = 2 \left|  M_{12}^q \right|$~\cite{DiLuzio:2019jyq}.
The new phases $\phi_q^\Delta$ are constrained by the measurements of the mixing phases $\sin 2\beta $ and $\sin 2 \beta_s$ in the golden plated modes
$B_d \to J/\psi K_S$ and $B_s \to J/\psi \phi$ to be at most
of the order of 1 or 2 degrees 
(except if one is willing to allow fine-tuned cancellations between new physics in $B$ mixing and penguin diagrams contributing to the $b \to c \bar{c} s$ decay).
Therefore in the case of new physics only acting in $M_{12}$,  the potential sizes of $a_{\rm fs}^q$ could be of the order of $10^{-4}$.
This is considerably below the current experimental accuracy, and the possible enhancement is not large enough to allow for an unambiguous observation at LHCb with $300\,{\rm fb}^{-1}$.

\subsection{\boldmath Modification of $\Gamma_{12}$}
\label{sec:Mod_G12}

The absorptive part of $B$ mixing is in general affected by new physics as (in the convention of \cite{Lenz:2011zz}) 
\begin{equation}
    \Gamma_{12}^q
    ~=~ 
    \Gamma_{12}^{q, \rm SM} \cdot \tilde{\Delta}_q
    ~=~ 
    \left| \Gamma_{12}^{q, \rm SM} \right| \cdot \left| \tilde{\Delta}_q
    \right| e^{i (\phi_q^{\Gamma, \rm SM} - \phi_q^{\tilde{\Delta}})} \, .
\end{equation}
In this case we get constraints from the measurements of the decay rate differences  $\Delta \Gamma_q$
\begin{eqnarray}
\Delta \Gamma_q 
& = & 
2 \left| \Gamma_{12}^{q} \right| \cos (\phi_{12}^q ) 
~=~ 2 \left| \Gamma_{12}^{q, \rm SM } \right|
\cdot \left| \tilde{\Delta}_q    \right|
\cos (\phi_{12}^{q, \rm SM } + \phi_q^\Delta +  \phi_q^{\tilde{\Delta}}) \, .
\end{eqnarray}
For $\Delta \Gamma_s$, experimental measurements~\cite{HFLAV:2019otj,LHCb:2019nin,ATLAS:2020lbz,CMS:2020efq} agree well with theory~\cite{Lenz:2019lvd} with a relative theory precision of the order of $15 \%$.
This translates into a maximal size of the new phase $\phi_s^{\tilde{\Delta}}$ of the order of $30^\circ$.
There could also be some further, less pronounced, enhancement due to modifications in~$| \tilde{\Delta}_q |$.
Such a sizable new phase $\phi_s^{\tilde{\Delta}}$ would lead to a strong enhancement of $a_{\rm fs}^s$, close to the current experimental bound, since 
\begin{eqnarray}
a_{\rm fs}^q &= &\frac{|\Gamma_{12}^q|  }{|M_{12}^q|  } \sin   \phi_{12}^q 
~=~ a_{\rm fs}^{q, \rm SM} \frac{\left| \tilde{\Delta}_q    \right|}{\left| {\Delta}_q    \right|} 
\frac{\sin (\phi_{12}^{q, \rm SM } + \phi_q^\Delta +  \phi_q^{\tilde{\Delta}}) }{\sin   \phi_{12}^{q, \rm SM} } 
\; .
\end{eqnarray}
There is even more space for a possible enhancement of $a_{\rm fs}^d$ via beyond SM (BSM) effects in $\Gamma_{12}^d$ (see also Refs.~\cite{Gershon:2010wx,Bobeth:2014rda}), since there are only relatively weak experimental constraints on $\Delta \Gamma_d $~\cite{HFLAV:2019otj,ATLAS:2016mln}.
This strongly motivates improved experimental measurements of $a_{\rm fs}^s$ and $a_{\rm fs}^d$. 

\subsection{\boldmath Modification of the $ B \to f$ decay amplitude}
As mentioned earlier, measurements of the rates of colour-allowed non-leptonic decays seem to deviate significantly from SM predictions~\cite{Huber:2016xod,Bordone:2020gao,Cai:2021mlt,Endo:2021ifc}.
Ref.~\cite{Bordone:2020gao} quotes for the decay $\bar{B}_s \to D_s^+ \pi^- $ a deviation
of the measurement from the QCD factorisation prediction of about four standard deviations.
In the case of the CKM suppressed decay $\bar{B}_d \to D^+ K^-$ this deviation is even larger than five standard deviations.
Commonly CKM leading, non-leptonic tree-level decays have been considered to be insensitive to new physics effects. However,
general bounds on BSM effects in non-leptonic tree-level decays were systematically studied in 
Refs.~\cite{Bobeth:2014rda,Brod:2014bfa,Lenz:2019lvd}, with results revealing that there is a sizable allowed parameter space for new effects, which do not violate any theoretical or experimental bound.
More recently such effects have also been investigated for the case of the decay $\bar{B}_s \to D_s^+ K^-$~ \cite{Fleischer:2021cct,Fleischer:2021cwb}. 
BSM explanations have been considered in \cite{Cai:2021mlt,Iguro:2020ndk} and challenged by collider bounds in \cite{Bordone:2021cca}.

Within the SM the decays $\bar{B}_s \to D_s^+ \pi^-$ and $\bar{B}_d \to D^+ K^-$ are
flavour-specific and CP conserving. 
Thus, using these decay to determine the asymmetries $A_{\rm fs}^q$, we expect to get the tiny values $a_{\rm fs}^q$.
However, if BSM effects modify the decay amplitudes, 
the relation between $A_{\rm fs}^q$ and $a_{\rm fs}^q$ of Eq.~\eqref{ea:def_asl3} is altered. 
Under the presence of general new physics contributions the decay amplitude of either $B_s \to D_s^- \pi^+$ or $B_d \to D^- K^+$ can be written as
\begin{eqnarray}
{\cal A}_f &= & \left| {\cal A}_f^{\rm SM} \right|  e^{i \phi^{\rm SM}}  e^{i \varphi^{\rm SM}}
+
\left| {\cal A}_f^{\rm BSM} \right|  e^{i \phi^{\rm BSM}}  e^{i \varphi^{\rm BSM}} 
\nonumber
\\
 &=: & \left| {\cal A}_f^{\rm SM} \right|  e^{i \phi^{\rm SM}}  e^{i \varphi^{\rm SM}}
\left(
1 +
r   e^{i \phi}  e^{i \varphi}
\right) \, , 
\label{amplitudeGeneral}
\end{eqnarray}
with relative strong $\phi = \phi^{\rm BSM} - \phi^{\rm SM}$ 
and weak $\varphi = \varphi^{\rm BSM} - \varphi^{\rm SM}$ phases,
and $r = | {\cal A}_f^{\rm BSM}|/| {\cal A}_f^{\rm SM}|$.
The amplitude $\bar{\cal A}_{\bar f}$ for the CP conjugate process is identical to ${\cal A}_f $ up to a change in the sign of
$\varphi$.
This allows now for direct CP violation in these decays,
challenging condition {\bf C2}. 
Nonetheless, the decays are expected to remain flavour specific, since we do not see a realistic possibility to sizably violate condition {\bf C1}: e.g.\ at the quark level the decay $\bar{B}_s \to D_s^+ \pi^-$ looks like
$b \bar{s} \to c \bar{s} \bar{u} d$, while a decay into the CP conjugate final state,
triggered by an $b \bar{s} \to s \bar{c} \bar{d} u$ quark level transition would require at least dimension-nine six-quark operators.%
\footnote{Condition {\bf C1} is also challenging to test experimentally, although this has been considered in \cite{Williams:1974106}.} 
Inserting
\begin{eqnarray}
\left| {\cal A}_f \right|^2 &= &  
\left| {\cal A}_f^{\rm SM} \right|^2  
\left[
1 + r^2 + 
2 r  (  \cos \phi  \cos \varphi - \sin \phi  \sin \varphi)
\right] \, ,
\nonumber
\\
\left| \bar{\cal A}_{\bar{f}} \right|^2 &= &  \left| {\cal A}_f^{\rm SM} \right| ^2 
\left[
1 + r^2 +
2 r  (  \cos \phi  \cos \varphi + \sin \phi  \sin \varphi)
\right] \, ,
\label{amplitudeGeneral2}
\end{eqnarray}
into Eq.~\eqref{ea:def_asl2} leads to
\begin{equation}
A_{\rm fs}^q ~=~ 
\frac{  a_{\rm  fs}^q - 2 r \sin \phi  \sin \varphi
+ 2 a_{\rm  fs}^q r \cos \phi  \cos \varphi  +  a_{\rm  fs}^q r^2 }
     {1 + 2  r \cos \phi  \cos \varphi  +  r^2 - 2 a_{\rm  fs}^q r \sin \phi  \sin \varphi}
     ~\approx~
      a_{\rm fs}^q - A^q_{\rm dir} \, ,
\label{ea:def_asl4}
\end{equation}
with the direct CP asymmetry 
$A^q_{\rm dir} \approx 2 r \sin \phi  \sin \varphi$ 
(formally defined in Appendix~\ref{app:DCP}, Eq.~\eqref{ea:def_other2}).\footnote{
Note that $a_{\rm fs}^q$ is defined as an asymmetry between the final states $f$ and $\bar{f}$, while $A^q_{\rm dir}$ is defined as an asymmetry between $\bar{f}$ and $f$, hence they appear with different signs in Eq.~\eqref{ea:def_asl4}.
}
To obtain the last expression in Eq.~(\ref{ea:def_asl4}) we have assumed $ a_{\rm  fs}^q$  and $ r$
to be small quantities and we have expanded up to leading order in these small parameters.
Allowing now for a size of $r \approx 0.1$, which is indicated by the studies in \cite{Bordone:2020gao,Huber:2016xod,Cai:2021mlt,Endo:2021ifc},
one can get -- depending on the values of the phases $\phi$ and $ \varphi$ -- values  of up to $|A_{\rm  fs}^q| = 0.2$, which are several orders of magnitude larger 
than the SM values of $a_{\rm  fs}^q$.

Thus, if the experimental value for $A_{\rm fs}^s(D_s^+ \pi^-)$ or $A_{\rm fs}^d(D^+ K^-)$ differs significantly from zero, with the currently achievable experimental precision, one has an unambiguous BSM signal, independent of any theory uncertainties.
Moreover, the effects of BSM contributions in $M_{12}$ and $\Gamma_{12}$, which affect $a_{\rm fs}^s$, can be separated from those in the decay amplitude, which affect $A_{\rm  fs}^s$, if we make the assumption that there is no direct CP violation in semileptonic decays 
which holds to excellent accuracy within the SM (since only one decay amplitude is contributing)
and to some extent also beyond the SM~\cite{Bar-Shalom:2010pkp,Descotes-Genon:2012oaf}.
In this case $A_{\rm  fs}^s(D_s^+ \pi^-) - A_{\rm  fs}^s(D_s^+ \ell^- \bar \nu_{\ell})$ gives a clean determination of  $A_{\rm dir}^s (D_s^+ \pi^-)$, and likewise $A_{\rm  fs}^d(D^+ K^-) - A_{\rm  fs}^d(D^+ \ell^- \bar \nu_{\ell}) = -A_{\rm dir}^d (D^+ K^-)$. 

Neither $A_{\rm  fs}^s(D_s^+ \pi^-)$ nor $A_{\rm  fs}^d(D^+ K^-)$ has yet been experimentally measured.
It is, however, likely that any large asymmetry in $\bar{B}_s \to D_s^+ \pi^-$ decays would have been spotted as this mode has been used for precise determinations of the $B_s$ oscillation frequency~\cite{LHCb:2021moh} and lifetime~\cite{LHCb:2014wet}, as well as being a control channel for CP violation studies in  $\bar{B}_s \to \Dspm \Kmp$ decays~\cite{LHCb:2017hkl}.
In what follows we focus on the $\bar{B}_s \to D_s^+ \pi^-$ mode as this appears to have the potential for precise measurements, but experimental studies of CP violation in $\bar{B}_d \to D^+ K^-$ decays are also well motivated.

\section{Untagged CP asymmetry}
In an Appendix, we present several further possible CP asymmetries that can be determined with flavour-specific decays, that have contributions from direct CP violation and/or CP violation in mixing. 
In that respect we will need, in addition to Eq.~\eqref{GammabarBf2} and~\eqref{GammaBbarf2}, the decay-rate evolution for neutral $B_q$ mesons that decay with the same flavour to that at production.
Assuming condition {\bf C1} is satisfied, these rates are given by~\cite{osti_1151538,Artuso:2015swg} 
\begin{eqnarray}
\Gamma \left[\bar{B}_q (t) \to \bar{f} \right] 
& = & 
\frac{1}{2} N_f \left|\bar{\cal A}_{\bar f}\right|^2
e^{-\Gamma_q t}
\label{GammabarBbarf2}
X_q^+(t)
\; ,
\\
\Gamma \left[B_q (t) \to f \right] 
& = & 
\frac{1}{2}
N_f \left|{\cal A}_f\right|^2 e^{-\Gamma_q t}
\label{GammaBf2}
X_q^+(t)
\; .
\end{eqnarray}
A particularly interesting observable is the untagged CP asymmetry, $A^q_{\rm untagged}$, given by 
\begin{equation}
A^q_{\rm untagged} 
= 
\frac{\left[\Gamma (\bar {B}_q (t)  \to \bar{f}) + \Gamma ( {B}_q (t)  \to \bar{f})\right] - \left[\Gamma (\bar {B}_q (t)  \to f) + \Gamma ({B}_q (t)  \to f)\right]}
{\left[\Gamma (\bar {B}_q (t)  \to \bar{f}) + \Gamma ( {B}_q (t)  \to \bar{f})\right] + \left[\Gamma (\bar {B}_q (t)  \to f) + \Gamma ({B}_q (t)  \to f)\right]} \, .
\label{ea:th-untagged}
\end{equation}
Inserting
Eq.~\eqref{GammabarBf2}, \eqref{GammaBbarf2}, \eqref{GammabarBbarf2} and
\eqref{GammaBf2}, we obtain
\begin{eqnarray}
A^q_{\rm untagged}
& = &
\frac{
\left|\bar{\cal A}_{\bar f}\right|^2
\left[X_q^+(t)
+
(1 - {a}_{\rm fs}^q)X_q^-(t)\right]
-
\left|{\cal A}_{f}\right|^2
\left[X_q^+(t)
+
(1 + {a}_{\rm fs}^q) X_q^-(t) \right]
}{\left|\bar{\cal A}_{\bar f}\right|^2
\left[X_q^+(t)
+
(1 - {a}_{\rm fs}^q) X_q^-(t) \right]
+
\left|{\cal A}_{f}\right|^2
\left[X_q^+(t)
+
(1 + {a}_{\rm fs}^q) X_q^-(t) \right]} 
\label{ea:eval-untagged-ext}
\\
& = & 
\frac{2 r \sin \phi  \sin \varphi -  {a}_{\rm fs}^q \left( 1 + 2 r \cos \phi  \cos \varphi + r^2 \right)  {Y}(t) }{1 + 2 r \cos \phi  \cos \varphi +  r^2 - 2 {a}_{\rm fs}^q r  \sin \phi  \sin \varphi \, {Y}(t) }
 \, ,
\end{eqnarray}
with 
\begin{eqnarray}
{Y}(t) & = & \frac{X_q^-(t)}{X_q^+(t)+X_q^-(t)}  
~=~ \frac{1}{2} 
\left[ 1 - \frac{\cos \left(\Delta M_q t \right)}{\cosh \left(\frac{\Delta \Gamma_q \, t}{2}\right)} \right].
\end{eqnarray}
Neglecting CP violation in mixing, ${a}_{\rm fs}^q = 0$, we find
\begin{equation}
A^q_{\rm untagged} ~=~ 
\frac{  2 r \sin \phi  \sin \varphi }
{1 + 2  r \cos \phi  \cos \varphi  +  r^2 }          
~=~ A^q_{\rm dir}
\, ,
\label{ea:eval-untagged}
\end{equation}
while  neglecting direct CP violation would give
 \begin{eqnarray}
A^q_{\rm untagged} & = & 
-{a}_{\rm fs}^q   {Y}(t) \, .
\end{eqnarray}
Generally, expanding everything up to linear terms in $r$ and ${a}_{\rm fs}^q$, we get 
\begin{eqnarray}\label{eq:TD-Auntagged}
A^q_{\rm untagged} & \approx &
A^q_{\rm dir} - {a}_{\rm fs}^q   {Y}(t) \, .
\end{eqnarray}
In contrast to Eq.~\eqref{ea:def_asl2}, this asymmetry is not independent of time.
It is, however, a convenient approach with which to study $B^0$ decays since it allows different sources of asymmetry to be disentangled. 
Measurements of $a_{\rm fs}^d$ have been made by fitting this time-dependent untagged asymmetry, using semileptonic decays in which the contribution from $A^d_{\rm dir}$ is expected to vanish~\cite{D0:2012idx,LHCb:2014dcj}.

For the $B_s$ case, it is experimentally convenient to measure the untagged asymmetry of time-integrated decay rates
\begin{eqnarray}\label{eq:Auntagged}
\hspace*{-5mm}
\langle A^q_{\rm untagged} \rangle
& \! = \! &
\frac{\int_0^\infty d t \left[\Gamma (\bar {B}_q (t)  \to \bar{f}) + \Gamma ( {B}_q (t)  \to \bar{f})\right] 
- 
\int_0^\infty d t 
\left[\Gamma (\bar {B}_q (t)  \to f) + \Gamma ({B}_q (t)  \to f)\right]}
{\int_0^\infty d t \left[\Gamma (\bar {B}_q (t)  \to \bar{f}) + \Gamma ( {B}_q (t)  \to \bar{f})\right] 
+ 
\int_0^\infty d t 
\left[\Gamma (\bar {B}_q (t)  \to f) + \Gamma ({B}_q (t)  \to f)\right]} 
\\
 & \! = \! & 
\frac{4 r \sin \phi  \sin \varphi -  {a}_{\rm fs}^q (1 - \rho_q) 
\left(1 + 2 r \cos \phi  \cos \varphi + r^2 \right)}
{2 \, (1 + 2 r \cos \phi  \cos \varphi + r^2 - {a}_{\rm fs}^q (1 -\rho_q) 
r \sin \phi  \sin \varphi)}\,,
\end{eqnarray}
where
\begin{equation}
\rho_q ~=~ \frac{\Gamma_q^2 - \frac{\Delta \Gamma_q^2}{4}}{\Gamma_q^2 + \Delta M_q^2}\, \quad
\left( \rho_d \approx 0.63 \ \text{and} \ \rho_s \approx 0.001 \right) \, .
\label{eq:rhoq}
\end{equation}
Expanding again up to linear terms in $r$ and ${a}_{\rm fs}^q$
one obtains:
\begin{equation}\label{eq:Auntagged2}
\langle A^q_{\rm untagged} \rangle 
~\approx~ 
A^q_{\rm dir} - \frac{{a}_{\rm fs}^q}{2} (1 - \rho_q).
\end{equation}
In the case of $B_s$ decays, where the oscillation frequency is fast compared to the lifetime, the dilution factor multiplying ${a}_{\rm fs}^q$ is effectively only 0.5.
Since determining $\langle A^q_{\rm untagged} \rangle$ avoids the need to tag the flavour of the $B_s$ meson at production, this is therefore an experimentally attractive approach with which to measure $a_{\rm fs}^s$, as quantified below.
This been exploited in existing measurements with semileptonic decays where the $A^q_{\rm dir}$ term is assumed to be zero~\cite{D0:2012hft,LHCb:2016ssr,LHCb:2013sxc}.
The same approach is also used for measurements of direct CP violation in modes where the $a_{\rm fs}^q$ contribution is negligible, for example $\bar{B}^0 \to K^-\pi^+$ and $\bar{B}_s \to K^+\pi^-$~\cite{LHCb:2020byh}.
In this case, the use of the untagged asymmetry does not cause any dilution of the sensitivity to $A^q_{\rm dir}$.
Note, however, that if $a_{\rm fs}^d$ or $a_{\rm fs}^s$ were as large in magnitude as $5\times10^{-3}$, at the extreme of their currently experimentally allowed ranges, this would according to Eq.~\eqref{eq:Auntagged2} induce a correction of about $1~(2.5) \times 10^{-3}$ in every $A_{\rm dir}$ measurement made with untagged $B^0$~($B_s$) decays.

We now consider the experimental prospects for measurements of $\langle A^s_{\rm untagged} \rangle$ in $\bar{B}_s \to D_s^+ \pi^-$ decays.
The LHCb experiment appears to have by far the best prospects to determine this quantity precisely, having previously demonstrated the capability to obtain large, low-background, samples in this decay channel~\cite{LHCb:2021moh}.
In addition to the existing data sample, corresponding to $9\,\invfb$ of $pp$ collision data collected in Runs~1 and~2 of the Large Hadron Collider, an additional $\approx 15\,\invfb$ of data is anticipated to be recorded during Run~3 with an upgraded detector~\cite{LHCbCollaboration:2319756}. 
A new, fully software-implemented, trigger strategy that will be utilised during Run~3 means that LHCb will benefit from enhanced efficiency for hadronic decay modes such as $\bar{B}_s \to D_s^+ \pi^-$. 

Based on the yields available in the existing data~\cite{LHCb:2021moh}, and the increase anticipated to be forthcoming with Run~3, we project a sensitivity to $\langle A^s_{\rm untagged} \rangle$ in $\bar{B}_s \to D_s^+ \pi^-$ decays of $\mathcal{O}(10^{-3})$. 
If systematic uncertainties can be controlled, it will be possible to further reduce this uncertainty as a total sample of up to $300 \,\invfb$ is collected by LHCb through operation in subsequent LHC run periods~\cite{LHCb:2018roe}.
As discussed above new physics contributions to tree-level amplitudes may modify this value from its tiny SM value to  $\mathcal{O}(10^{-2})$ or above, and hence the experimental measurement will either discover or significantly constrain these BSM effects. 
Measurements of ${A}_{\rm fs}^s$ with semileptonic decays are expected to be even more precise, and will constrain the contribution to $\langle A^s_{\rm untagged} \rangle$ from ${a}_{\rm fs}^s$, assuming no direct CP violation in semileptonic decays. 
Indeed, the existing limits on ${a}_{\rm fs}^s$ from semileptonic measurements, which are consistent with the tiny SM expectation, are sufficient to conclude that a non-zero value of $\langle A^s_{\rm untagged} \rangle$ in $\bar{B}_s \to D_s^+ \pi^-$ decays at the $\mathcal{O}(10^{-2})$ level would be clear evidence for BSM effects causing direct CP violation.

Experimentally, the quantity that is directly measured is 
\begin{equation}
A_{\rm raw} ~=~ 
\frac{N(D_s^+\pi^-) - N(D_s^-\pi^+)}
{N(D_s^+\pi^-) + N(D_s^-\pi^+)}
\nonumber
\end{equation}
where $N(X)$ is the total number of $\Bs \rightarrow X$ and $\Bsb \rightarrow X$ decays observed in the data. 
This is related to $\langle A^s_{\rm untagged} \rangle$ by
\begin{equation}
\langle A^s_{\rm untagged} \rangle ~=~
A_{\rm raw} - A_{\rm det} - A_{\rm prod}\frac{\int_{t=0}^\infty e^{-\Gamma_s t} \cos(\Delta M_s t) \epsilon(t) dt}{\int_{t=0}^\infty e^{-\Gamma_s t} \cosh(\frac{\Delta \Gamma_s t}{2}) \epsilon(t) dt} - \sum_i f_{\rm bkg}^i \cdot A_{\rm bkg}^i \, .
\label{ea:exp-full}
\end{equation}
The detector asymmetries,
$A_{\rm det}$, will be reduced by reconstructing the \Dspm meson in the $\Dspm \rightarrow \Pphi \pipm$ final state. 
The $\bar{B}_s$ decay is then fully reconstructed in the symmetric $\Kpm \Kmp \pipm \pimp$ final state with the two kaons having approximately the same momentum distribution. 
A small detection asymmetry will remain due to the momentum difference between the \pipm originating from a $\bar{B}_s$ versus a \Dspm decay, but these effects can be understood using control samples~\cite{Veronesi-thesis}.  
(Similarly, if reconstruction and detection asymmetries of $\Kpm$ mesons are well understood, the whole Dalitz plot of $\Dspm \to \Kp\Km\pipm$ decays can be used to increase the available sample size.)

The $B^0_s$--$\bar{B}^0_s$ production asymmetry in $pp$ collisions with decays within the LHCb detector acceptance is denoted by $A_{\rm prod}$, and can be measured using the decay-time dependence of flavour-specific decays~\cite{LHCb:2017bdt}. 
Due to fast $B^0_s$--$\bar{B}^0_s$ oscillations, the impact of $A_{\rm prod}$ is significantly diluted by the time integral ratio in Eq.~\eqref{ea:exp-full}.
The tiny residual contribution can nonetheless be calculated and corrected for.
This calculation must also take into account the fact that the acceptance $\epsilon(t)$ depends on the $B$ meson decay time, and hence enters the integrals in Eq.~\eqref{ea:exp-full}.
(For completeness, the decay-time acceptance function should also be taken into account when determining Eq.~\eqref{eq:Auntagged}, which impacts on the dilution factor of Eq.~\eqref{eq:Auntagged2}.)

The asymmetries from various sources of background decays are accounted for through $A_{\rm bkg}^i$ which is the asymmetry of background contribution $i$. 
Each background contribution is given a weight, $f_{\rm bkg}^i$, according to its relative fraction in the data.
Since the background fractions are low, the sources of background are well-understood~\cite{LHCb:2021moh} and their asymmetries can be determined from control samples, this is not expected to provide a limiting systematic uncertainty. 

As previously noted, by not attempting to distinguish between the mixed $\bar{B}_s(t) \to D_s^- \pi^+$ decays and the unmixed $\bar{B}_s(t) \to D_s^+ \pi^-$ decays, there is a significant gain in the statistics available to measure $A^s_{\rm dir}$.
This is much greater than the factor of 2 one would naively expect from the large value of $\Delta M_s$, since one no longer requires initial state flavour tagging.
LHCb has achieved a tagging efficiency for $\bar{B}_s$ mesons of $\epsilon_{\rm tag} \approx 80\%$ and a mistag rate of $w\approx 36\%$~\cite{LHCb:2021moh}, giving an effective tagging efficiency of $\epsilon_{\rm tag} \left(1-2w\right)^2 \approx 6\%$. 
Consequently, untagged methods are highly preferable for the studies of CP asymmetries in $B_s$ mesons discussed here.

To determine $A_{\rm dir}^d$ for the flavour-specific $B^0 \to D^+K^-$ decays, on the other hand, it would be preferable to study the decay-time dependence of the untagged asymmetry as given in Eq.~\eqref{eq:TD-Auntagged}.
Once experimental effects are taken into account it can be shown that fitting this distribution allows the separate measurement of the combinations $A_{\rm dir}^d + A_{\rm det} + a_{\rm fs}^d/2$ and $A_{\rm prod} + a_{\rm fs}^d/2$~\cite{LHCb:2014dcj,LHCb:2017bdt}.
Hence it is necessary to take as an external input the value of $a_{\rm fs}^d$ obtained from semileptonic decays (under the assumption of no direct CP violation).  
The detection asymmetry $A_{\rm det}$ can be determined from control samples as before, although in this case with the favoured $\Dpm \to \Kmp\pipm\pipm$ decay the final state is not symmetric so one cannot benefit from cancellations of asymmetries as in the case of $\bar{B}_s \to \Dsp\pim$.

\section{Conclusion}
We have studied the CP asymmetries that can be investigated using flavour-specific decays, with particular attention to the non-leptonic decays $\bar{B}_s \to D_s^+ \pi^-$ and $\bar{B}^0 \to D^+ K^-$ that have not previously been used for this purpose. 
Within the SM no direct CP violation occurs in these decays, and they be used to determine the flavour-specific CP asymmetry $a_{\rm fs}^q$, albeit with worse precision than obtained with semileptonic decays.
If new physics appears only in $B$ mixing then semileptonic decays will still be superior in the experimental determination of the flavour-specific CP asymmetry.
This changes, however, as soon as new CP violating contributions to the non-leptonic decays are allowed.
In this case the tiny effects due to $a_{\rm fs}^q$ might be completely overshadowed by the contributions stemming from direct CP violation.

Experimentally the $\bar{B}_s \to D_s^+ \pi^-$ decay is particularly attractive, due to the large available yield, the symmetric final state, and the fact that measurements can be made without the need to determine the production flavour of the \Bs mesons in $\bar{B}_s \to D_s^+ \pi^-$. 
The untagged and time-integrated CP asymmetries depend on both $a_{\rm fs}^s$ and direct CP violation and are therefore sensitive to BSM effects in either. 
We expect a sensitivity of around one per mille for the untagged asymmetry in $\bar{B}_s \to D_s^+ \pi^-$ decays can be achieved at LHCb with Run~3 data, with improvement possible as larger data samples are collected further into the future.
This will allow BSM effects causing direct CP violation in these decays to either be discovered or significantly constrained.
Once the precision reaches a level that is sensitive to the Standard Model value of $a_{\rm fs}^s$, one can consider the difference in $A_{\rm fs}^s$ values measured in $\bar{B}_s \to D_s^+ \pi^-$ and in semileptonic decays.
Any significantly non-zero value of this difference would be an unambiguous signal of new physics, not relying on any theoretical estimates of non-perturbative contributions. 

\section*{Acknowledgment}
The work of AL and AR was supported by the BMBF project 05H21PSCLA: 
„Verbundprojekt 05H2021 (ErUM-FSP T04) - Run 3 von LHCb am LHC:
Theoretische Methoden für LHCb und Belle II“. The work of NS is supported by the ERC Starting Grant 852642 - Beauty2Charm.

\appendix

\section{Appendix}
Here for completeness we present expressions for other CP asymmetries.
\subsection{Direct CP asymmetry}\label{app:DCP}
The direct CP asymmetry can be defined as the asymmetry of the decay-rates for neutral $B_q$ mesons that decay with the same flavour as at production (see Eqs.~\eqref{GammabarBbarf2} and~\eqref{GammaBf2}), i.e.\  
\begin{eqnarray}
A_{\rm dir}^q & = & \frac{\Gamma \left( \bar{B}_q (t)  \to \bar{f}\right) 
- \Gamma \left({B}_q (t)  \to f\right) }
 {\Gamma \left(\bar{B}_q (t)  \to \bar{f}\right) + \Gamma \left({B}_q (t)  \to f\right) } 
~=~
\frac{ \left|\bar{\cal A}_{\bar f}\right|^2 - \left|{\cal A}_f\right|^2 
  }
{ \left|\bar{\cal A}_{\bar f}\right|^2 +\left|{\cal A}_f\right|^2 
  }
~=~ 
\frac{  2 r \sin \phi  \sin \varphi
 }
{1 + 2  r \cos \phi  \cos \varphi  +  r^2 } \,,
\nonumber
\\
&\approx & 
 2 r \sin \phi  \sin \varphi 
\, \,,
\label{ea:def_other2}
\end{eqnarray}
where the approximation is a good one for $r \ll 1$.
It is simply the asymmetry of the decay amplitudes squared, and hence also equal to the asymmetry of decay rates at $t=0$, and to the untagged CP asymmetry in the limit of negligible ${a}_{\rm fs}^q$.
\subsection{Indirect CP asymmetry}
Indirect CP asymmetry is typically defined as
\begin{eqnarray}
A_{\rm ind}^q 
& = &
\frac{\Gamma \left( \bar{B}_q (t)  \to f \right) 
- \Gamma \left({B}_q (t) \to f \right)}
{\Gamma \left(\bar{B}_q (t) \to {f}\right) 
+ \Gamma \left({B}_q (t)  \to f \right)} 
=
-
\frac{
 2 \cos  \left( \Delta M_q t\right)
\, - {a}_{\rm fs}^q \, 
X_q^-(t)
}
{
 2 \cosh \left( \frac{\Delta \Gamma_q t}{2} \right) + 
 {a}_{\rm fs}^q 
 X_q^-(t)
 } 
\, .
\label{ea:def_CPother4}
\end{eqnarray}
Since the definition of this asymmetry involves only one final state, for flavour-specific decays it does not depend on $A^q_{\rm dir}$.
The dominant contribution to this asymmetry is given by
$ \cos  \left( \Delta M_q t\right) / \cosh \left( \Delta \Gamma_q t/2 \right)$, with a small correction proportional to ${a}_{\rm fs}^q$.
\\
In a similar way one may also define 
\begin{eqnarray}
\tilde A_{\rm ind}^q 
& = &
\frac{\Gamma \left(\bar{B}_q (t)  \to \bar f \right) 
- \Gamma \left({B}_q (t)  \to \bar f \right) }
{\Gamma \left(\bar{B}_q (t)  \to \bar f \right) 
+ \Gamma \left({B}_q (t)  \to \bar f \right) } 
=
\frac{
2 \cos  \left( \Delta M_q t\right)
+ {a}_{\rm fs}^q 
X_q^-(t)
}
{
 2 \cosh \left( \frac{\Delta \Gamma_q t}{2} \right) -
 {a}_{\rm fs}^q 
 X_q^-(t)
 } 
\, ,
\label{ea:def_CPother4-tilde}
\end{eqnarray}
which gives up to an overall sign the same result as $A_{\rm ind}^q $, when 
$a_{\rm fs}^q$ is replaced by $- a_{\rm fs}^q$.
Note that if we add these two asymmetries, then the leading terms cancel and the sum is proportional to $a_{\rm fs}^q$:
\begin{eqnarray}
 A_{\rm ind}^q + \tilde A_{\rm ind}^q 
& \approx &
 {a}_{\rm fs}^q 
\left[
1 - \frac{\cos^2  \left( \Delta M_q t\right)}
{\cosh^2 \left( \Delta \Gamma_q t/2 \right)}
\right]
\, .
\end{eqnarray}
This provides a possibility to determine ${a}_{\rm fs}^q$ independently of, and with no assumption on, $A^q_{\rm dir}$.

In addition, we consider the time-integrated indirect 
$CP$-asymmetries
\begin{eqnarray}
\langle A_{\rm ind}^q \rangle 
= 
\frac{\int_0^\infty d t \left[\Gamma (\bar{B}_q (t)  \to f) - \Gamma ({B}_q (t)  \to f) \right]}
{\int_0^\infty d t \left[\Gamma (\bar{B}_q (t)  \to {f}) + \Gamma ({B}_q (t)  \to f) \right]} \, ,
\\[2mm]
\langle \tilde A_{\rm ind}^q \rangle 
= 
\frac{\int_0^\infty d t \left[\Gamma (\bar{B}_q (t)  \to \bar f) - \Gamma ({B}_q (t)  \to \bar f) \right]}
{\int_0^\infty d t \left[\Gamma (\bar{B}_q (t)  \to \bar f) + \Gamma ({B}_q (t) \to \bar f) \right]} \, . 
\end{eqnarray}
Using Eqs.~\eqref{GammabarBf2}, \eqref{GammaBbarf2},
\eqref{GammabarBbarf2} and  \eqref{GammaBf2} we obtain
\begin{eqnarray}
 \langle A_{\rm ind}^q \rangle 
 =
- \frac{\rho_q - a^q_{\rm fs} \, R_q}{1 + a^q_{\rm fs} \, R_q} \, , 
\qquad 
\langle \tilde A_{\rm ind}^q \rangle 
 =
\frac{\rho_q + a^q_{\rm fs} \, R_q}{1 - a^q_{\rm fs} \, R_q} \, ,
\end{eqnarray}
where $\rho_q$ is defined in Eq.~\eqref{eq:rhoq} and
\begin{equation}
R_q = \frac{\frac{\Delta \Gamma_q^2}{4} + \Delta M_q^2}{2 \, (\Gamma_q^2 + \Delta M_q^2)} \, .
\label{eq:Rhoq}
\end{equation}
The time-integrated asymmetries have a leading dependence on
$\rho_q$ and small corrections proportional to $a_{\rm fs}^q$.
Note, that this leading term cancels in the sum of 
the two time-integrated CP asymmetries:
\begin{equation}
\langle A_{\rm ind}^q \rangle + \langle \tilde A_{\rm ind}^q \rangle 
\approx
2 \, a^q_{\rm fs} \, R_q (1 + \rho_q) \, .
\end{equation}
Since $\Delta \Gamma_s \ll \Gamma_ s \ll \Delta M_s$, one can expand further to get
\begin{equation}
\langle A_{\rm ind}^s \rangle + \langle \tilde A_{\rm ind}^s \rangle 
\approx a_{\rm fs}^s \left (1 - \frac{\Gamma_s^4}{\Delta M_s^4} \right).
\end{equation}

\subsection{Mixed CP asymmetry}

We can also look at the  asymmetry
\begin{equation}
A_{\rm mix}^q 
= 
\frac{\Gamma ({B}_q (t)  \to \bar{f} ) - \Gamma ({B}_q (t)  \to f)}
{\Gamma ({B}_q (t)  \to \bar{f}) + \Gamma ({B}_q (t)  \to f) } \, ,
\label{ea:def_CPother6}
\end{equation}
where we get for flavour-specific decays
\begin{eqnarray}
A_{\rm mix}^q = 
\frac{
\left|\bar{\cal A}_{\bar f}\right|^2
 (1 - {a}_{\rm fs}^q) X_q^- (t)
- 
\left|{\cal A}_f\right|^2
X_q^+ (t)
}
{
\left|\bar{\cal A}_{\bar f} \right|^2
 (1 - {a}_{\rm fs}^q ) \, X^-_q (t) 
 + \left|{\cal A}_f\right|^2
X_q^+ (t)
} 
= - \frac{F_q (t) - f(r, a_{\rm fs}^q,\phi, \varphi)}
{1 -  f(r,a_{\rm fs}^q,\phi, \varphi) F_q (t) } \, ,
\label{ea:def_CPother5} 
\end{eqnarray}
with the auxiliary functions 
\begin{eqnarray}
F_q (t)
& = & 
\frac{\cos  \left( \Delta M_q t\right)}
{\cosh \left( \frac{\Delta \Gamma_q t}{2} \right)} \, ,
\\
f(r,a,\phi, \varphi) & = & \frac{4 r  \sin \phi  \sin \varphi - a - 2 a r (\sin \phi  \sin \varphi+ \cos \phi  \cos \varphi) - a r^2}
{ 2 + 4 r \cos \phi  \cos \varphi - a + 2 r^2 - 2 a r (\sin \phi  \sin \varphi+ \cos \phi  \cos \varphi) - a r^2} \, .
\label{eq:f}
\end{eqnarray}
Keeping only terms linear in $a_{\rm fs}^q$ and $r$, we arrive at
\begin{eqnarray}
f(r,a_{\rm fs}^q,\phi, \varphi) & \approx & 
A^q_{\rm dir} - \frac{a_{\rm fs}^q}{2}  \, ,
\end{eqnarray}
with $A^q_{\rm dir} $ given in Eq.~\eqref{ea:def_other2}.
Having no new physics in the decay $\bar{B}_s \to D_s^+ \pi^-$,
we get $f(0,a_{\rm fs}^s,\phi, \varphi) = -a_{\rm fs}^s/(2 - a_{\rm fs}^s) \approx -a_{\rm fs}^s/2$, while
for a sizable phases $\phi$ and $\varphi$ and for larger
values of $r$ we can neglect~$a_{\rm fs}^s$ and get 
$f(r,0,\phi, \varphi) \approx A_{\rm dir}^q $.
In the approximation of keeping only the linear terms in $a_{\rm fs}^q$ and $r$ the asymmetry looks as
\begin{eqnarray}
A_{\rm mix}^q  & \approx & 
-
\frac{\cos  \left( \Delta M_q t\right)}
{\cosh \left( \frac{\Delta \Gamma_q t}{2} \right)}
+ 
 \left[
  A^q_{\rm dir} - \frac{a_{\rm fs}^q}{2} 
 \right]
 \left[ 1 -\frac{\cos^2  \left( \Delta M_q t\right)}
{\cosh^2 \left( \frac{\Delta \Gamma_q t}{2} \right)}\right] \, . 
 \end{eqnarray}
As in the case of the indirect CP asymmetires the dominant contribution to this asymmetry is given by
$ \cos  \left( \Delta M_q t\right) / \cosh \left( \Delta \Gamma_q t/2 \right)$, 
but now the small corrections are  proportional to $ r$ (in $A_{\rm dir}^q$) 
and ${a}_{\rm fs}^q$.
\\
And again, one can define a similar asymmetry
\begin{equation}
\tilde A_{\rm mix}^q 
= 
\frac{\Gamma \left(\bar {B}_q (t)  \to \bar{f}\right) 
- \Gamma \left(\bar {B}_q (t)  \to f \right)}
{\Gamma \left(\bar {B}_q (t)  \to \bar{f} \right) 
+ \Gamma \left(\bar {B}_q (t)  \to f \right) } \, ,
\label{ea:def_CPother6-tilde}
\end{equation}
for which we get
\begin{eqnarray}
\tilde A_{\rm mix}^q 
& = & 
\frac{
\left|\bar{\cal A}_{\bar f}\right|^2
X_q^+ (t)
- 
\left|{\cal A}_f\right|^2
(1 + {a}_{\rm fs}^q)
X_q^- (t)
}
{
\left|\bar{\cal A}_{\bar f}\right|^2
X_q^+ (t) 
+ 
\left|{\cal A}_f\right|^2
(1 + {a}_{\rm fs}^q) 
X_q^- (t)
}
=
\frac{F_q (t) - f(r, -a_{\rm fs}^q, \phi, -\varphi)}
{1 -  f(r, -a_{\rm fs}^q, \phi, -\varphi) F_q (t) } 
\label{ea:def_CPother5-tilde}
\\
 & \approx & 
\frac{\cos  \left( \Delta M_q t\right)}
{\cosh \left( \frac{\Delta \Gamma_q t}{2} \right)}
+ 
 \left[
 A^q_{\rm dir} - \frac{a_{\rm fs}^q}{2} 
 \right]
 \left[ 1 -\frac{\cos^2  \left( \Delta M_q t\right)}
{\cosh^2 \left( \frac{\Delta \Gamma_q t}{2} \right)}\right] \, . 
\end{eqnarray}
One can get rid of the dominant contributions in 
 $A_{\rm mix}^q$ and $ \tilde A_{\rm mix}^q$ by considering the sum of the two, to obtain
 an observable that is directly proportional to
 $2 r  \sin \phi  \sin \varphi - a_{\rm fs}^q/2$:
\begin{eqnarray}
 A_{\rm mix}^q  + \tilde A_{\rm mix}^q 
 & \approx & 
2
 \left[
 A^q_{\rm dir} - \frac{a_{\rm fs}^q}{2} 
 \right]
 \left[ 1 -\frac{\cos^2  \left( \Delta M_q t\right)}
{\cosh^2 \left( \frac{\Delta \Gamma_q t}{2} \right)}\right] \, . 
\end{eqnarray}
Defining the time-integrated mixed $CP$-asymmetries
\begin{eqnarray}
\langle A_{\rm mix}^q \rangle 
= 
\frac{\int_0^\infty d t \left[\Gamma (B_q (t)  \to \bar f) - \Gamma ({B}_q (t)  \to f) \right]}
{\int_0^\infty d t \left[\Gamma (B_q (t)  \to \bar f) + \Gamma ({B}_q (t)  \to f) \right]},
\\[2mm]
\langle \tilde A_{\rm mix}^q \rangle 
= 
\frac{\int_0^\infty d t \left[\Gamma (\bar{B}_q (t)  \to \bar f) - \Gamma (\bar{B}_q (t)  \to \bar f) \right]}
{\int_0^\infty d t \left[\Gamma (\bar{B}_q (t)  \to \bar f) + \Gamma (\bar {B}_q (t) \to \bar f) \right]},
\end{eqnarray}
one obtains
\begin{eqnarray}
\langle A_{\rm mix}^q \rangle 
& = & 
- \frac{\rho_q - f(r,a_{\rm fs}^q,\phi, \varphi)}{1 -  f(r,a_{\rm fs}^q,\phi, \varphi) \rho_q},
\\
\langle \tilde A_{\rm mix}^q \rangle 
& = & 
\frac{\rho_q - f(r, - a_{\rm fs}^q, \phi, -\varphi)}{1 - f(r, -a_{\rm fs}^q, \phi, -\varphi) \rho_q }, 
\end{eqnarray}
where $f(r,a,\phi, \varphi) $ is given in Eq.~\eqref{eq:f}, and
$\rho_q$ in  Eq.~\eqref{eq:rhoq}.
For the sum of the time-integrated CP asymmetries we get:
\begin{eqnarray}
 \langle A_{\rm mix}^q \rangle  +  \langle \tilde A_{\rm mix}^q \rangle
 & \approx & 
 (2 A^q_{\rm dir} - a_{\rm fs}^q) (1 - \rho_q^2).
\end{eqnarray}

\bibliographystyle{JHEP}
\bibliography{References}

\end{document}